\documentclass{Interspeech}
\usepackage{amsmath,amssymb,amsfonts}
\usepackage{algorithmic}
\usepackage{graphicx}
\usepackage{textcomp}
\usepackage{xcolor}
\usepackage{float}
\usepackage{graphicx}
\usepackage{array}
\usepackage{float}
\usepackage{pgfplots}
\usepackage{placeins}

\interspeechcameraready

\title{Disentangling Dual-Encoder Masked Autoencoder for Respiratory Sound Classification}

\author[]{Peidong}{Wei}
\author[]{Shiyu}{Miao}
\author[]{Lin}{Li*}

\affiliation{School of Electronic Science and Engineering}{Xiamen University}{China}

\email{lilin@xmu.edu.cn}


\usepackage{comment}

\begin{document}

\maketitle

\begin{abstract}
    
Deep neural networks have been applied to audio spectrograms for respiratory sound classification, 
but it remains challenging to achieve satisfactory performance due to the scarcity of available data.
Moreover, domain mismatch may be introduced into the trained models as a result of the respiratory sound samples being collected from various electronic stethoscopes, patient demographics, and recording environments. To tackle this issue, we proposed a modified Masked Autoencoder (MAE) model, named Disentangling Dual-Encoder MAE (DDE-MAE) for respiratory sound classification. Two independent encoders were designed to capture disease-related and disease-irrelevant information separately, achieving feature disentanglement to reduce the domain mismatch. Our method achieves a competitive performance on the ICBHI dataset.
\end{abstract}

\section{Introduction}
Respiratory sound classification is an essential task in the early detection and diagnosis of respiratory diseases. Accurate classification can assist healthcare professionals in identifying conditions such as asthma, bronchitis, and pneumonia from respiratory sounds\cite{arabi2021covid}. However, several challenges arise due to the variability in recording conditions.

Respiratory sound data are collected from different patients using various stethoscopes in diverse environments. This variability introduces significant domain mismatch problems\cite{guan2021domain}, making it difficult for models to generalize across different datasets\cite{quinonero2022dataset}\cite{torralba2011unbiased}. Furthermore, in real-world scenarios, respiratory sound data are often from unseen domains, and obtaining information about these specific domains is frequently impractical. This makes the challenge of domain adaptation even more difficult to address.

Recently, numerous studies have attempted various methods to improve model performance. For instance, Audio Spectrogram Transformer with Patch-Mix and Contrastive Learning (AST + Patch-Mix CL)\cite{bae23b_interspeech} employs a mix-based data augmentation method and establishes a corresponding contrastive learning approach to address the issue of data scarcity. The Multi-View Spectrogram Transformer (MVST)\cite{he2024multi} adopts an ensemble approach that combines multiple models trained on different views of the data to achieve performance enhancement. Bridging Text and Sound Modalities (BTS)\cite{kim2024bts} is a multimodal approach, integrating textual information, such as patient demographics (age, gender) and stethoscope recording conditions, with audio data. Furthermore, some researchers have attempted to address domain adaptation by utilizing the classification labels of the stethoscope used during recording  and incorporating these labels into domain adversarial\cite{ganin2016domain}  training methods\cite{kim2024stethoscope}.
Although it is helpful in improving performance, the approach still heavily depends on the availability of labeled information, which is often limited. For instance, the ICBHI dataset \cite{rocha2018alpha} includes only four types of stethoscopes, and relying on these limited labels restricts the model's ability to generalize to new, unseen environments or stethoscope types.

\begin{figure*}[t]
    \centering
    \includegraphics[width=1\textwidth]{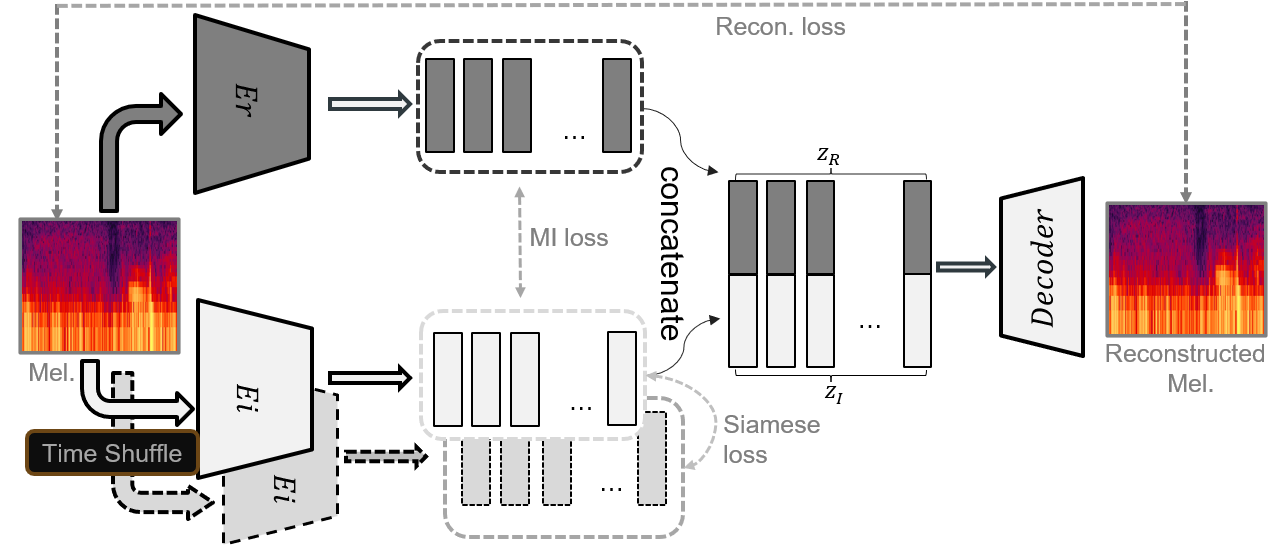}
    \caption{Overview of Proposed DDE-MAE}
    \label{fig:DDE-MAE}
\end{figure*}

In this paper, we propose a novel self-supervised approach that does not require label information of domain (e.g., types of stethoscopes). The overall structure, the Disentangling Dual-Encoder Masked Autoencoder\cite{he2022masked} (DDE-MAE) is shown in Fig.~\ref{fig:DDE-MAE}, which leverages the power of self-supervised learning to disentangle disease-related features from disease-irrelevant features. Hence, our model can achieve effective domain adaptation without the need for extensive domain labels.

Our DDE-MAE model employs two independent encoders. The disease-related encoder captures features directly related to respiratory diseases. The disease-irrelevant encoder captures background and non-disease-related features by processing both the original spectrogram and a time-shuffled version of the spectrogram, using a Siamese loss to enforce feature invariance to temporal changes. Additionally, To ensure that the embeddings captured by the two encoders are independent, we employ the variational Contrastive Log-ratio Upper Bound (vCLUB) algorithm \cite{pmlr-v119-cheng20b} to estimate and minimize the mutual information between the embeddings from the disease-related encoder and the disease-irrelevant encoder. 
Our contributions can be summarized as follows:
\begin{itemize}
\item We introduce a dual-encoder architecture for respiratory sound classification that disentangles disease-related and disease-irrelevant information.
\item We propose a self-supervised learning approach that addresses domain adaptation without relying on domain information.
\item We demonstrate the effectiveness of our method on the ICBHI dataset, achieving a competitive performance and highlighting its potential for real-world applications.
\end{itemize}

\section{Methodology}
\subsection{Overview of Proposed DDE-MAE}

Our proposed model, Disentangling Dual-Encoder MAE (DDE-MAE), leverages the principles of the Masked Autoencoder (MAE) architecture to address the domain adaptation challenges in respiratory sound classification. The MAE architecture typically consists of an encoder-decoder structure. The encoder processes the input spectrograms by masking a portion of the input and learning to represent the unmasked parts efficiently. The decoder reconstructs the masked parts from the encoded representations, ensuring that the encoder captures meaningful features of the input data\cite{he2022masked}.

To disentangle disease-related and disease-irrelevant information, we modify the MAE architecture to include two independent encoders:

\textbf{Disease-Related Encoder. }This encoder (denoted as \(E_r\)) captures features directly related to respiratory diseases by processing the input spectrograms without modifications.

\textbf{Disease-Irrelevant Encoder. }This encoder  (denoted as \(E_i\)) captures background and non-disease-related features. It processes two inputs in parallel: the original spectrogram and a time-shuffled version. The time-shuffling process involves randomly shuffling the time axis of the spectrogram, disrupting temporal dependencies while preserving spectral content. A Siamese loss function \cite{chicco2021siamese} enforces similarity between the representations of the shuffled and original spectrograms, encouraging the encoder to focus on static features.

The decoder architecture remains consistent with the original MAE design\cite{he2022masked}, combining the outputs from both encoders to reconstruct the masked portions of the input spectrograms. This reconstruction ensures that the encoders learn complementary features that together capture the full information content of the input. 

\subsection{vCLUB Algorithm}
\label{ssec:subhead}
To ensure the embeddings captured by the two encoders, \(Z_R\) and \(Z_I\), are independent, we utilize a mutual information (MI)\cite{shannon1948mathematical} loss. 
Specifically, we employ the variational Contrastive
Log-ratio Upper Bound (vCLUB) algorithm to estimate and minimize the mutual information between the embeddings from the disease-related encoder and the disease-irrelevant encoder.

Mutual information is a measure of the dependence between two different variables \(A\) and \(B\), formulated as:
\begin{equation}
\mathcal{I}(A, B)=\int_A \int_B P(A, B) \log \frac{P(A, B)}{P(A) P(B)}
\end{equation}
where \(P(A)\) and \(P(B)\) are the marginal distributions of \(A\) and
\(B\) respectively, and \( P(A, B)\) denotes the joint distribution of
\(A\) and \(B\). 

Without the exact distribution functions of the variables, computing the precise value of mutual information is usually challenging. The vCLUB algorithm provides an efficient method to estimate the upper bound of mutual information, which can be used to minimize mutual information. 
For two different variables \(A\) and \(B\), it is defined as:

\begin{equation}
\begin{aligned}
I(A, B)= & \mathbb{E}_{p(A, B)}\left[\log q_\theta(B \mid A)\right] \\
& - \mathbb{E}_{p(A)} \mathbb{E}_{p(B)}\left[\log q_\theta(B \mid A)\right]\label{eq}
\end{aligned}
\end{equation}
where \(q_\theta(B |A)\) is a variational distribution with parameter \(\theta\) to approximate \(p(B |A)\).

In DDE-MAE, the upper
bound of MI between the disease-related representation \(Z_R\) and the disease-relevant representation \(Z_I\) is derived as:

\begin{equation}
\begin{aligned}
I(Z_R, Z_I)= & \mathbb{E}_{p(Z_R, Z_I)}\left[\log q_\theta(Z_I \mid Z_R)\right] \\
& - \mathbb{E}_{p(Z_R)} \mathbb{E}_{p(Z_I)}\left[\log q_\theta(Z_I \mid Z_R)\right]
\end{aligned}
\end{equation}

Then, the estimator for vCLUB with sample \{\(z_{R_i}, z_{I_i}\)\} from \{\(Z_{R}, Z_{I}\)\} is:

\begin{align}
L_{MI}(z_{R_i}, z_{I_i}) &= \frac{1}{N^2} \sum_{i=1}^N \sum_{j=1}^N 
\left[
\log q_\theta\left(z_{I_i} \mid z_{R_i}\right) \right. \nonumber \\
&\quad \left. -\log q_\theta\left(z_{I_j} \mid z_{R_i}\right)
\right].
\end{align}

\subsection{Training Procedure}
\label{ssec:subhead}

We fine-tune the Audio-MAE \cite{huang2022masked} model (MAE pretrained on AudioSet\cite{gemmeke2017audio}) in two stages:

(i) Fine-Tuning for disentanglement: We duplicate the encoder to create the disease-related and disease-irrelevant encoders:

\textbf{Input Shuffling.} For each spectrogram \(X\), we generate a time-shuffled version \(\Tilde{X}\) and input both of them into the disease-irrelevant encoder in parallel.

\textbf{Masked Autoencoder Training.} Similar to pretraining, we still use the reconstruction loss to reconstruct the spectrogram, formulated by:

\begin{equation}
\mathcal{L}_{\text{recon}} = \frac{1}{N} \sum_{i=1}^{N} \lVert x_i - \hat{x}_i \rVert_2^2
\end{equation}
where $x_i$ is the sample of original spectrogram $X$, $\hat{x}_i$ is the sample of  reconstructed spectrogram $\hat{X}$ and $\lVert \cdot \rVert_2$ represents the L2 norm (Euclidean distance).

\textbf{Siamese Loss Application.} We apply the Siamese loss to the outputs of the disease-irrelevant encoder, formulated by: 
\begin{equation}
\mathcal{L}_{\text{siamese}} = \frac{1}{N} \sum_{i=1}^{N} \max(\lVert x_i - \tilde{x}_i\rVert_2 - \mathit{margin}, 0)
\end{equation}
where $\lVert \cdot \rVert_2$ represents the distance computed using the Euclidean distance, \(\mathit{margin}\) is a hyperparameter that defines the threshold for the loss, and \(x_i\) and \(\Tilde{x_i}\) are samples from  \(X\) and \(\Tilde{X}\), respectively.

\textbf{Mutual Information Loss Application.} We apply the MI loss using the vCLUB algorithm to ensure the embeddings from both encoders are independent.

The loss of the disentanglement stage can be computed as:

\begin{equation}
\label{total loss}
\mathcal{L} = \mathcal{L}_{\text{recon}} + \alpha_1 \mathcal{L}_{\text{siamese}} + \alpha_2 \mathcal{L}_{\text{MI}}
\end{equation}
where \(\alpha_1\) and \(\alpha_2\) are weight factors for siamese loss and MI loss.

(ii) Fine-Tuning for Classification: After disentanglement training, we fine-tune the disease-related encoder for the classification task on the ICBHI dataset using cross-entropy loss.

\section{Experiments}
\subsection{Dataset}
The proposed DDE-MAE is evaluated on the largest publicly available respiratory sound dataset, the ICBHI dataset\cite{rocha2018alpha},
which contains 6,898 respiratory cycles, including 3,642 normal breathing, 1,864 crackling breathing, 886 wheezing breathing, and 506 breathing with both crackling and wheezing. 
The sampling rates are 4kHz, 10kHz, or 44.1kHz, with recording durations between 10 and 90 seconds. To ensure consistency in processing, all recordings were downsampled to 4kHz before further analysis. Each recording includes an average of seven breathing cycles. 
Following the official experimental settings, 60$\%$ of the recordings are randomly selected for training, and the remaining 40$\%$  are used for testing. 
\subsection{Metrics}
The classification performance is evaluated using a metric called \textit{Score}, which is the average of specificity (\( S_p \)) and sensitivity (\( S_e \)). These two metrics are defined as follows:

\begin{equation}
    S_p = \frac{C_n}{N_n}
    \end{equation}
\begin{equation}
\quad S_e = \frac{C_c + C_w + C_b}{N_c + N_w + N_b}
    \end{equation}

In these expressions, \( C_i \) and \( N_i \) represent the number of correctly classified samples and the total number of samples for each class \( i \in \{n, c, w, b\} \) corresponding to \{\text{normal}, \text{crackle}, \text{wheeze}, \text{both}\}, respectively.

\subsection{Traning details}

For the disentanglement stage, we set the batch size to 8 and the initial learning rate to \(2 \times 10^{-4}\). Additionally, we empirically set \(\alpha_1 = 0.5\) and \(\alpha_2 = 0.02\) in equation (\ref{total loss}) to facilitate model convergence. In the classification training stage, we train the model using a cross-entropy loss function with an initial learning rate of \(1 \times 10^{-3}\) for 50 epochs. The AdamW optimizer \cite{Loshchilov2017DecoupledWD} with a fixed weight decay of \(5 \times 10^{-4}\) is used in both stages.

\subsection{Ablation Study}
An ablation study is conducted to evaluate the effectiveness of the proposed DDE-MAE. We firstly explore on the ICBHI dataset:

\textbf{Baseline 1 (B1): Direct Training for classification.} The MAE model is directly trained on the classification task using the ICBHI dataset.


\textbf{System 1 (S1): Training for disentanglement and classification.} The MAE model undergoes disentanglement training with the dual-encoder setup before being trained on the classification task using the ICBHI dataset.


To further investigate the effectiveness of the pretrained model, e.g. pretraining on AudioSet, we introduce three additional configurations, mirroring the above settings but with AudioSet pertaining:

\textbf{Baseline 2 (B2): Fine-tuning for classification}. The MAE pretrained on AudioSet is directly fine-tuned on the classification task using the ICBHI dataset.

\textbf{System 2 (S2): Fine-tuning for reconstruction and classification.} The MAE model pretrained on AudioSet, undergoes further fine-tuning on the ICBHI dataset through a reconstruction task before fine-tuning on the classification task.

\textbf{System 3 (S3): Fine-Tuning for disentanglement and classification.}
The MAE pretrained on AudioSet is fine-tuned by disentanglement training with the dual-encoder setup, and then on the classification task using the ICBHI dataset. 




\begin{table*}
    \caption{Comparison of Different Training Methods with Confidence Intervals}
    \begin{center}
    \small  
    \resizebox{\textwidth}{!}{  
    \begin{tabular}{@{}cccccc@{}}
    \hline
    No. & Model & Method & SP(\%) & SE(\%) & AS(\%) \\ 
    \hline
    B1 & MAE (Encoder only) & Direct Training & \textbf{78.21\tiny$\pm$1.53} & 21.24\tiny$\pm$0.75 & 49.73\tiny$\pm$0.92 \\
    S1 & DDE-MAE & Training for disentanglement and classification & 77.34\tiny$\pm$1.45 & \textbf{26.23\tiny$\pm$1.01} & \textbf{51.78\tiny$\pm$0.89} \\

    \hline
    \hline
    B2 & MAE (Encoder only) & Fine-tuning for classification & 64.47\tiny$\pm$1.25 & 50.72\tiny$\pm$0.68 & 57.59\tiny$\pm$0.91 \\
    S2 & MAE & Fine-tuning for reconstruction and classification & 68.89\tiny$\pm$1.32 & 50.98\tiny$\pm$0.76 & 59.93\tiny$\pm$0.79 \\
    S3 & DDE-MAE & Fine-tuning for reconstruction and classification & \textbf{69.32\tiny$\pm$1.27} & \textbf{53.69\tiny$\pm$1.15} & \textbf{61.50\tiny$\pm$0.94} \\

    \hline
    \end{tabular}
    }
    \end{center}
    \label{tab:combined}
\end{table*}

The results are shown in Table~\ref{tab:combined}, which demonstrate that our method outperforms the baselines in both scenarios: with and without pretraining on AudioSet. Specifically, S1 shows an improvement over B1 with a 5.01\% increase in SE and a 2.05\% increase in AS. Similarly, S3 shows significant improvements over B2, with a 4.85\% increase in SP, a 2.97\% increase in SE, and a 3.91\% increase in AS. Besides, the comparisons between B1 and B2, as well as between S1 and S3 reveal that pretraining plays a pivotal role in enhancing model performance across all evaluated metrics.

The inclusion of the ``Fine-tuning for reconstruction and classification'' (S2) configuration is critical to discerning the true impact of disentanglement training. It allows us to determine whether the performance improvements are due to the disentanglement structure itself or simply because the model is exposed to ICBHI data during additional training. It is shown that S3 outperforms S2 in both SP and SE, validating the effectiveness of the disentanglement mechanism in enhancing model performance.


In addition, to evaluate the effectiveness of our disentanglement training approach, we conducted an ablation study comparing four different configurations.

The results are 
summarized in Table~\ref{tab:ablation3}. It can be observed that the model with both MI loss and Siamese loss (Complete DDE-MAE) outperforms the other configurations in terms of SP, SE, and AS. This highlights the importance of both the MI loss and the Siamese loss in capturing and separating disease-related and disease-irrelevant features.
\begin{table}[H]
\centering
\caption{Ablation study results comparing the Complete DDE-MAE model with different loss configurations.}
\begin{tabular}{lccc}
\hline
Method & SP(\%) & SE(\%) & AS(\%) \\ 
\hline
DDE-MAE & \textbf{69.32\tiny$\pm$1.27} & \textbf{53.69\tiny$\pm$1.15} & \textbf{61.50\tiny$\pm$0.94} \\
- MI Loss & 69.11\tiny$\pm$1.20 & 53.50\tiny$\pm$1.10 & 61.30\tiny$\pm$0.85 \\
- Siamese Loss & 68.54\tiny$\pm$1.15 & 51.74\tiny$\pm$1.02 & 60.14\tiny$\pm$0.88 \\
\hspace{1em}- MI Loss & 67.61\tiny$\pm$1.08 & 48.71\tiny$\pm$0.98 & 58.16\tiny$\pm$0.91 \\ 
\hline
\end{tabular}
\label{tab:ablation3}
\end{table}
\FloatBarrier
\subsection{Qualitative Analysis}

\begin{figure}[H]
    \centering
    \begin{tikzpicture}
        \begin{axis}[
            ybar,
            bar width=12pt, 
            width=8cm, 
            height=6cm, 
            enlarge x limits=0.8, 
            ymin=0, ymax=100,
            xtick=data, 
            xticklabels={Acc. on devices(\%), Acc. on patients(\%)}, 
            symbolic x coords={Acc. on devices(\%), Acc. on patients(\%)}, 
            nodes near coords, 
            legend style={at={(0.5,-0.15)}, anchor=north, legend columns=-1}, 
            tick label style={font=\scriptsize}, 
            legend style={
                at={(0.9,0.95)}, 
                anchor=north, 
                legend columns=1, 
                font=\scriptsize 
            },
        ]
        \addplot[
            style={blue,fill=blue!30},
            bar shift=-14pt 
        ] coordinates {(Acc. on devices(\%),81.5) (Acc. on patients(\%),70.3)};
        
        \addplot[
            style={red,fill=red!30},
            bar shift=10pt 
        ] coordinates {(Acc. on devices(\%),93.2) (Acc. on patients(\%),85.6)};
        
        \legend{$E_R$, $E_I$} 
        \end{axis}
    \end{tikzpicture}
    \caption{Analysis of Encoder Performance on Device and Patient Classification}
    \label{fig:analysis}
\end{figure}
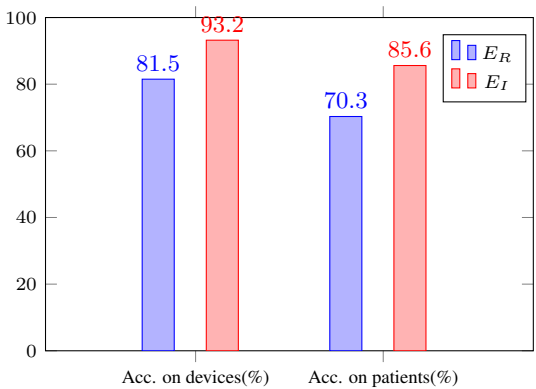

To evaluate the effectiveness of the disentanglement, we incorporated a linear classification head into each encoder and conducted classification training using cross-entropy loss on device and patient labels, as illustrated in Fig.~\ref{fig:analysis}. This analysis aimed to assess the ability of each encoder to capture domain-specific features. The results demonstrated that the disease-irrelevant encoder \(E_I\) effectively isolated and captured features specific to the device and patient, as expected. On the other hand, the disease-related encoder \(E_R\) was less influenced by these device- and patient-specific factors and primarily focused on disease-specific characteristics. However, it was not entirely immune to these extraneous factors, showing a slight degree of influence, which suggests that further refinement may be necessary to fully disentangle these influences. This experiment highlights the effectiveness of the disentanglement strategy while also revealing some remaining challenges in fully isolating disease-related information.

\subsection{Comparison to State-of-the-art Methods}
Many deep neural methods have been utilized for respiratory sound classification, such as LungBRN\cite{ma2019lungbrn}, LungAttn\cite{li2021lungattn}, ResNeSt\cite{wang2022domain}, RespireNet\cite{gairola2021respirenet}, ARSCNet\cite{xu2021arsc}, CNN6\cite{moummad2023pretraining}, Co-tuned ResNet-50\cite{nguyen2022lung}, audio spectrogram transformer with patch-mix and contrastive learning (AST + patch-mix CL)\cite{bae23b_interspeech}, stethoscope-guided supervised contrastive learning(SG-SCL)\cite{kim2024stethoscope}. The comparison with these state-of-the-art methods is summarized in Table ~\ref{tab:sota}. It can be observed that the proposed DDE-MAE demonstrates competitive performance among models with similar parameter counts, achieving the highest sensitivity, which is a crucial metric for detecting abnormal respiratory sounds. 

In addition, some approaches that rely on model ensembles \cite{he2024multi}, mixup \cite{DBLP:conf/iclr/ZhangCDL18} or cutmix \cite{yun2019cutmix} data augmentation methods \cite{bae23b_interspeech}, additional modality information \cite{kim2024bts}, or extra domain labels \cite{kim2024stethoscope}, our method does not require such additional techniques or information.
\vspace{-0.5em}
\begin{table}[H]
\centering
\caption{Comparisons with state-of-the-art methods}
\label{tab:sota}
\scriptsize 
\begin{tabular}{ccccc}
\hline
   Method  & SP(\%) & SE(\%) & AS(\%) \\ 
\hline
LungBRN\cite{ma2019lungbrn} & 69.20 & 31.10 & 50.16 \\
LungAttn\cite{li2021lungattn} & 71.44 & 36.36 & 53.90 \\
ResNeSt\cite{wang2022domain} & 70.40 & 40.20 & 55.30 \\
RespireNet\cite{gairola2021respirenet} & 72.30 & 40.10 & 56.20 \\
ARSC-Net\cite{xu2021arsc} & 67.13 & 46.38 & 56.76 \\
CNN6\cite{moummad2023pretraining} & 75.95 & 39.15 & 57.55 \\
Co-tuned ResNet-50\cite{nguyen2022lung} & 79.34 & 37.24 & 58.29 \\
AST + patch-mix CL\cite{bae23b_interspeech} & \textbf{81.66} & 43.01 &\textbf{62.37}\\
SG-SCL\cite{kim2024stethoscope} &79.87& 43.55 &61.71\\
DDE-MAE (ours) &69.32\tiny$\pm$1.27 & \textbf{53.69\tiny$\pm$1.15} & 61.50\tiny$\pm$0.94 \\
\hline
\end{tabular}
\end{table}


   
\section{Conclusion}
In this work, we propose the Disentangling Dual-Encoder Masked Autoencoder (DDE-MAE) for respiratory sound classification, addressing domain adaptation challenges due to data variability from different patients, stethoscopes, and environments. By employing two encoders—one capturing disease-related features and the other capturing disease-irrelevant features using Siamese loss and Mutual Information (MI) loss with the vCLUB algorithm—we ensure independent and focused feature extraction. This disentangling approach improves the model's robustness by effectively separating task-relevant information from confounding factors, enabling better generalization across diverse scenarios. Our self-supervised approach, validated on the ICBHI dataset, achieved competitive performance, demonstrating the effectiveness of disentangling relevant and irrelevant information for robust disease classification in real-world applications.
\section{Acknowledgements}
Thanks to the National Natural Science Foundation of China(Grant No.62371407 and No.62001405) for funding.
\newpage

\bibliographystyle{IEEEtran}
\bibliography{mybib}

\end{document}